\documentclass[english,technote,onecolumn,12pt]{IEEEtran}
\usepackage{lmodern}
\usepackage[T1]{fontenc}
\usepackage[latin9]{inputenc}
\usepackage{color}
\usepackage{babel}
\usepackage{array}
\usepackage{float}
\usepackage{booktabs}
\usepackage{units}
\usepackage{multirow}
\usepackage{amsmath}
\usepackage{graphicx}
\usepackage[unicode=true,pdfusetitle,
 bookmarks=true,bookmarksnumbered=false,bookmarksopen=false,
 breaklinks=false,pdfborder={0 0 0},pdfborderstyle={},backref=false,colorlinks=true]
 {hyperref}
\hypersetup{
 linkcolor=blue,citecolor=blue,urlcolor=blue}

\makeatletter

\providecommand{\tabularnewline}{\\}

\usepackage[font=normalsize]{caption}
\usepackage[noadjust]{cite}
\newenvironment{cellvarwidth}[1][t]
    {\begin{varwidth}[#1]{\linewidth}}
    {\@finalstrut\@arstrutbox\end{varwidth}}
\usepackage{varwidth}

\makeatother

\begin{document}
\title{Electromagnetic Field Near The Common Edge of a Perfectly Conducting
Wedge and a Resistive Half-plane}
\author{Igor M. Braver, Pinkhos Sh. Fridberg, Khona L. Garb, Iosif M. Yakover}
\maketitle
\begin{abstract}
The behavior of the electromagnetic field near a common edge of a
resistive half-plane and a perfectly conducting wedge is investigated.
The possible appearance besides power terms also of logarithmic functions
in the field expansions at specific angles between a resistive plane
and the sides of the wedge is indicated. It is shown that in the main
terms of the electric or magnetic field components perpendicular to
the edge, logarithmic functions appear only in two particular cases:
when a half-plane forms a straight angle with one of the sides of
the conducting wedge, or when the external angle of the conducting
wedge is a right one. Otherwise, the main terms of all components
have a purely power character and can be determined by the Meixner
method. To confirm the efficiency of accounting for the found field
distribution law, transversal wave numbers of the dominant mode of
a rectangular waveguide with a resistive film in the diagonal plane
is calculated.
\end{abstract}

\section{Problem statement}

\begin{figure}

\begin{centering}
\includegraphics{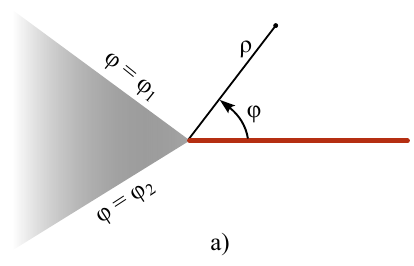}\includegraphics{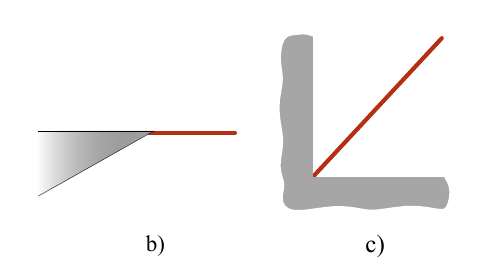}
\par\end{centering}
\caption{\label{fig:1}Perfectly conducting wedges and resistive films.}

\end{figure}

Let us consider a structure consisting of a two-dimensional perfectly
conducting wedge and a resistive half-plane (RHP).The wedge and the
RHP are assumed to have a common edge (Fig.~\ref{fig:1}a). In formulating
the boundary value problems of electrodynamics describing structures
with sharp edges, the uniqueness of the solution can be established
only if the electromagnetic energy stored in any finite neighborhood
of the edge that does not contain sources is finite. Using this condition,
we determine the behavior of the electromagnetic field $\{\mathbf{E},\mathbf{H}\}$
and the surface current density $\mathbf{j}$ on the RHP near the
edge.

The field $\{\mathbf{E},\mathbf{H}\}$ in the source-free region satisfies
the homogeneous Maxwell equations
\begin{equation}
\nabla\times\mathbf{E}=ik_{0}\zeta_{0}\mathbf{H},\qquad\nabla\times\mathbf{H}=-i\frac{k_{0}}{\zeta_{0}}\mathbf{E},
\end{equation}
and the following boundary conditions:
\begin{align}
 & E_{u}\Bigr|_{\varphi=\varphi_{\nu}}=E_{\rho}\Bigr|_{\varphi=\varphi_{\nu}}=0,\qquad\nu=1;2,\nonumber \\
 & E_{u}\Bigr|_{\varphi=0}=E_{\rho}\Bigr|_{\varphi=2\pi}=Wj_{u},\qquad j_{u}=-\left(H_{\rho}\Bigr|_{\varphi=0}-H_{\rho}\Bigr|_{\varphi=2\pi}\right),\label{eq:BC}\\
 & E_{\rho}\Bigr|_{\varphi=0}=E_{u}\Bigr|_{\varphi=2\pi}=Wj_{\rho},\qquad j_{\rho}=H_{u}\Bigr|_{\varphi=0}-H_{u}\Bigr|_{\varphi=2\pi}.\nonumber 
\end{align}
Here $k_{0}$ and $\zeta_{0}$ are wave number and wave impedances
of the free space; $\rho$, $\varphi$, $u$ are cylindrical coordinates;
$W$ is surface impedance. We will assume that the dependence of the
field on the coordinate $u$ is given by the function $I(u)=I_{0}\exp(ih_{0}u)$,
where $I_{0}$ is a dimensional factor. Let us introduce into consideration
two dimensionless scalar functions $\psi$ and $\chi$ proportional
to the longitudinal (relative to the edge) components of the field:
\begin{equation}
H_{u}(\rho,\varphi,u)=I(u)\psi(\rho,\varphi),\qquad E_{u}(\rho,\varphi,u)=\zeta_{0}I(u)\chi(\rho,\varphi).
\end{equation}
Functions $\psi$ and $\chi$ satisfy the two-dimensional Helmholtz
equation

\begin{equation}
\left[\frac{1}{r}\frac{\partial}{\partial r}\left(r\frac{\partial}{\partial r}\right)+\frac{1}{r^{2}}\frac{\partial^{2}}{\partial\varphi^{2}}+1\right]\begin{Bmatrix}\psi\\
\chi
\end{Bmatrix}=0,\label{eq:Helmholtz}
\end{equation}
where $r=\kappa\rho$, $\kappa=\sqrt{k_{0}^{2}-h_{0}^{2}}$. Expressing
the transverse components of the field in terms of the longitudinal
ones and using (\ref{eq:BC}), we obtain the boundary conditions for
$\psi$ and $\chi$:
\begin{align}
 & \chi\Bigr|_{\varphi=\varphi_{\nu}}=\frac{\partial\psi}{\partial\varphi}\Bigr|_{\varphi=\varphi_{\nu}}=0,\qquad\nu=1;2,\nonumber \\
 & \chi\Bigr|_{\varphi=0}=\chi\Bigr|_{\varphi=2\pi}=-w\left[\left.\left(ih\frac{\partial\psi}{\partial r}-\frac{ik}{r}\frac{\partial\chi}{\partial\varphi}\right)\right|_{\varphi=0}-\left.\left(ih\frac{\partial\psi}{\partial r}-\frac{ik}{r}\frac{\partial\chi}{\partial\varphi}\right)\right|_{\varphi=2\pi}\right],\label{eq:BC-psi-chi}\\
 & \frac{\partial\psi}{\partial\varphi}\Bigr|_{\varphi=0}=\frac{\partial\psi}{\partial\varphi}\Bigr|_{\varphi=2\pi},\quad\left.\left(\frac{ik}{r}\frac{\partial\psi}{\partial\varphi}+ih\frac{\partial\chi}{\partial r}\right)\right|_{\varphi=0}=w\left(\psi\Bigr|_{\varphi=0}-\psi\Bigr|_{\varphi=2\pi}\right),\quad k=\frac{k_{0}}{\kappa},\quad h=\frac{h_{0}}{\kappa},\quad w=\frac{W}{\zeta_{0}}.\nonumber 
\end{align}
Our further goal is to study the behavior of functions $\psi$ and
$\chi$ as $r\to0$.

\section{Field expansion near the edge}

Following the method of work \cite{Braver1986}, we seek the solution
of equation (\ref{eq:Helmholtz}) in the form

\begin{equation}
\begin{split}\psi & =r^{\tau}\sum_{m=0}^{\infty}\sum_{n=0}^{m}a_{mn}(\varphi)r^{m}\ln^{n}r,\\
\chi & =r^{\tau}\sum_{m=0}^{\infty}\sum_{n=0}^{m}b_{mn}(\varphi)r^{m}\ln^{n}r,
\end{split}
\quad\tau\geq0.\label{eq:psi-chi}
\end{equation}
Substituting (\ref{eq:psi-chi}) into (\ref{eq:Helmholtz}) and (\ref{eq:BC-psi-chi}),
we obtain that the sought for coefficients $a_{mn}(\varphi)$, $b_{mn}(\varphi)$
satisfy the infinite system of differential equations 

\begin{equation}
\begin{split} & a_{mn}^{\prime\prime}+(m+\tau)^{2}a_{mn}+2(m+\tau)(n+1)a_{m,n+1}\\
 & +(n+2)(n+1)a_{m,n+2}+a_{m-2,n}=0,\\
 & b_{mn}^{\prime\prime}+(m+\tau)^{2}b_{mn}+2(m+\tau)(n+1)b_{m,n+1}\\
 & +(n+2)(n+1)b_{m,n+2}+b_{m-2,n}=0,
\end{split}
\label{eq:a''b''}
\end{equation}
and boundary conditions
\begin{equation}
\begin{split} & b_{mn}(\varphi_{\nu})=a_{mn}^{\prime}(\varphi_{\nu})=0,\quad\nu=1;2\\
 & [h(m+\tau)a_{mn}(0)+h(n+1)a_{m,n+1}(0)-kb_{mn}^{\prime}(0)]\\
 & -[h(m+\tau)a_{mn}(2\pi)+h(n+1)a_{m,n+1}(2\pi)-kb_{mn}^{\prime}(2\pi)]=\frac{i}{w}b_{m-1,n}(0),
\end{split}
\label{eq:bc-1}
\end{equation}
\begin{equation}
\begin{split} & b_{mn}(0)-b_{mn}(2\pi)=0\\
 & a_{mn}^{\prime}(0)-a_{mn}^{\prime}(2\pi)=0\\
 & ka_{mn}^{\prime}(0)+h(m+\tau)b_{mn}(0)+h(n+1)b_{m,n+1}(0)\\
 & =-iw[a_{m-1,n}(0)-a_{m-1,n}(2\pi)].
\end{split}
\label{eq:bc-2}
\end{equation}
Here $a_{mn},b_{mn}\equiv0$ if $m<n$; prime denotes differentiation
with respect to $\varphi$. As can be seen from (\ref{eq:a''b''})--(\ref{eq:bc-2}),
the differential equations and boundary conditions for the diagonal
($m=n$) coefficients turn up to be homogeneous. Solution of these
equations that satisfy (\ref{eq:bc-1}) can be written in the form
\begin{equation}
\begin{split}a_{mm}(\varphi) & =P_{mm}^{(\nu)}\cos[(m+\tau)(\varphi-\varphi_{\nu})],\\
b_{mm}(\varphi) & =Q_{mm}^{(\nu)}\sin[(m+\tau)(\varphi-\varphi_{\nu})],
\end{split}
\label{eq:ab-solutions}
\end{equation}
where $\nu=1$ for region $0\leq\varphi\leq\varphi_{1}$ and $\nu=2$
for $\varphi_{2}\leq\varphi\leq2\pi$; $P_{mm}^{(\nu)}$, $Q_{mm}^{(\nu)}$
are constants of integration. Substituting (\ref{eq:ab-solutions})
into the boundary conditions (\ref{eq:bc-2}), we obtain a homogeneous
system of linear algebraic equations (SLAE) for determining $P_{mm}^{(\nu)}$,
$Q_{mm}^{(\nu)}$. Determinant of this system is 
\begin{equation}
\Delta_{m}=(m+\tau)^{3}\sin[(m+\tau)\alpha_{1}]\sin[(m+\tau)\alpha_{2}]\sin[(m+\tau)\alpha_{3}],
\end{equation}
where $\alpha_{1}=\varphi_{1}$, $\alpha_{2}=2\pi-\varphi_{2}$, $\alpha_{3}=\alpha_{1}+\alpha_{2}$.

From the condition of existence of a non-trivial solution of a homogeneous
SLAE for the integration constants $P_{00}^{(\nu)}$, $Q_{00}^{(\nu)}$
appearing in the main coefficients $a_{00}$, $b_{00}$, we obtain
a characteristic equation for determining $\tau$:
\begin{equation}
\Delta_{0}\equiv\tau^{3}\sin(\tau\alpha_{1})\sin(\tau\alpha_{2})\sin(\tau\alpha_{3})=0.\label{eq:Delta0}
\end{equation}
Non-negative solutions of this equation will be numbered in the increasing
order:
\begin{equation}
\tau_{1}=0,\quad\tau_{2}=\frac{\pi}{\alpha_{3}},\quad\tau_{3}=\frac{\pi}{\alpha_{1}}\quad\tau_{4}=\frac{2\pi}{\alpha_{3}},\ \ldots
\end{equation}
Hereafter without loss of generality we assume $\alpha_{1}\geq\alpha_{2}$.
For each of the found $\tau_{s}$, $s=1,2,3,\ldots$, there is a specific
set of $P_{00}^{(\nu)}$, $Q_{00}^{(\nu)}$ that do not turn into
zero all simultaneously, but are expressed in the terms of arbitrary
constants $A$, $B$, $C$, $D$, $\ldots$:
\begin{equation}
\begin{split} & P_{00}^{(\nu)}=A-(-1)^{\nu}B,\quad Q_{00}^{(\nu)}=0,\quad s=1;\\
 & Q_{00}^{(\nu)}=(-1)^{\nu+1}\frac{C}{\sin(\tau_{2}\alpha_{\nu})},\quad P_{00}^{(\nu)}=\frac{h}{k}Q_{00}^{(\nu)},\quad s=2;\\
 & P_{00}^{(\nu)}=D\delta_{1\nu},\quad Q_{00}^{(\nu)}=\frac{h}{k}P_{00}^{(\nu)},\quad s=3;\\
 & \cdots\cdots\cdots\cdots\cdots\cdots\cdots\cdots\cdots\cdots\cdots\cdots\cdots\cdots\cdots\cdots\cdots\cdots,
\end{split}
\label{eq:PQ}
\end{equation}
where $\delta_{1\nu}$ is the Kronecker symbol.

The process of determining the following coefficients $a_{mn}$, $b_{mn}$
depends on the value of the determinants $\Delta_{m}$ for $m\geq1$.
Let us first consider the general case, when no two roots of equation
(\ref{eq:Delta0}) differ by an integer and, consequently, the determinants
$\Delta_{1}$, $\Delta_{2}$, $\Delta_{3}$, $\ldots$ do not vanish
when any $\tau_{s}$ is substituted into them. In this case, for $m\geq1$
the homogeneous SLAE for $P_{mm}^{(\nu)}$, $Q_{mm}^{(\nu)}$ have
only zero solutions and therefore $a_{mn}=b_{mn}\equiv0$. By successively
decreasing the index $n$, we similarly verify that the coefficients
$a_{mn}$, $b_{mn}$ turn out to be zero for all $n$, except $n=0$.
This means that in (\ref{eq:psi-chi}) all terms containing logarithmic
functions disappear, and the expansions of the fields are described
by the usual Meixner power series \cite{Meixner1972,Mittra1971}.
The coefficients $a_{m0}$, $b_{m0}$ for $m\geq1$ are found from
the corresponding inhomogeneous equations and are expressed in terms
of the constants $A$, $B$, $C$, $D$, $\ldots$ introduced in (\ref{eq:PQ}).
The presence of these undefined constants is due to the fact that
in the problem under consideration external sources are not specified.
Without specifying the structure of these sources, let us examine
the general solution, which is obtained by summing the expansions
(\ref{eq:psi-chi}) over $s$, corresponding to all values of $\tau_{s}$.
Let us present here the first few terms of such a solution: 
\begin{equation}
\begin{split}(-1)^{\nu}\psi & =(-1)^{\nu}A-B+\frac{2iwk}{\sin\alpha_{\nu}}B\cos(\varphi-\varphi_{\nu})r\\
 & +\left[\left(\frac{w^{2}k^{2}\sin\alpha_{3}}{\sin\alpha_{1}\sin\alpha_{2}\sin2\alpha_{\nu}}-\frac{h^{2}\sin2\alpha_{3-\nu}}{2\sin2\alpha_{3}}\right)B\cos2(\varphi-\varphi_{\nu})\right.\\
 & +\left.\frac{1}{4}B-\frac{1}{4}(-1)^{\nu}A\right]r^{2}+\ldots+Ch\left\{ -\frac{\cos[\tau_{2}(\varphi-\varphi_{\nu})]}{k\sin(\tau_{2}\alpha_{\nu})}r^{\tau_{2}}+\right.\\
 & +\left.\frac{i\sin[(1+\tau_{2})\alpha_{3-\nu}]}{w(1+\tau_{2})\sin\alpha_{3}}\cos[(\tau_{2}+1)(\varphi-\varphi_{\nu})]r^{\tau_{2}+1}+\ldots\right\} +\\
 & +D\left\{ \delta_{1\nu}\cos(\tau_{3}\varphi)r^{\tau_{3}}-\frac{ikw}{1+\tau_{3}}\frac{\cos[(\tau_{3}+1)(\varphi-\varphi_{\nu})]}{\sin[(\tau_{3}+1)\alpha_{\nu}]}r^{\tau_{3}+1}+\ldots\right\} +\ldots
\end{split}
\label{eq:psi}
\end{equation}
\begin{equation}
\begin{split}(-1)^{\nu}\chi & =\frac{2iwh}{\sin\alpha_{\nu}}B\sin(\varphi-\varphi_{\nu})r\\
 & +kh\left(\frac{w^{2}\sin\alpha_{3}}{\sin\alpha_{1}\sin\alpha_{2}\sin2\alpha_{\nu}}-\frac{\sin2\alpha_{3-\nu}}{2\sin2\alpha_{3}}\right)B\sin2(\varphi-\varphi_{\nu})r^{2}+\ldots\\
 & +C\left\{ -\frac{\sin[\tau_{2}(\varphi-\varphi_{\nu})]}{\sin(\alpha_{\nu}\tau_{2})}r^{\tau_{2}}\right.\\
 & +\left.\frac{ik\sin[(1+\tau_{2})\alpha_{3-\nu}]}{w(1+\tau_{2})\sin\alpha_{3}}\sin[(1+\tau_{2})(\varphi-\varphi_{\nu})]r^{\tau_{2}+1}+\ldots\right\} \\
 & +Dh\left\{ \frac{\delta_{1\nu}}{k}\sin(\tau_{3}\varphi)r^{\tau_{3}}-\frac{iw}{1+\tau_{3}}\frac{\sin[(\tau_{3}+1)(\varphi-\varphi_{\nu})]}{\sin[(1+\tau_{3})\alpha_{\nu}]}r^{\tau_{3}+1}+\ldots\right\} +\ldots
\end{split}
\end{equation}
The obtained expressions of the functions $\psi$ and $\chi$ allow
us to determine the behavior near the edge of all field components
and the surface current density on the RHP. To this end, we will represent
the desired components in the form
\begin{equation}
\begin{split}H_{u} & =(-1)^{\nu}I\{(-1)^{\nu}A-B+o(r^{0})\},\\
E_{u} & =(-1)^{\nu}I\zeta_{0}\left\{ \frac{2iwh}{\sin\alpha_{\nu}}B\sin(\varphi-\varphi_{\nu})r-\frac{\sin\tau_{2}(\varphi-\varphi_{\nu})}{\sin(\tau_{2}\alpha_{\nu})}Cr^{\tau_{2}}+o(r^{\min(1,\tau_{2})})\right\} ,\\
E_{\rho} & =(-1)^{\nu}I\zeta_{0}\left\{ \frac{2w}{\sin\alpha_{\nu}}B\sin(\varphi-\varphi_{\nu})-\frac{i}{k}\tau_{3}\delta_{1\nu}\sin(\tau_{3}\varphi)Dr^{\tau_{3}-1}+o(r^{\min(0,\tau_{3}-1)})\right\} ,\\
E_{\varphi} & =(-1)^{\nu}I\zeta_{0}\left\{ \frac{2w}{\sin\alpha_{\nu}}B\cos(\varphi-\varphi_{\nu})-\frac{i}{k}\tau_{3}\delta_{1\nu}\cos(\tau_{3}\varphi)Dr^{\tau_{3}-1}+o(r^{\min(0,\tau_{3}-1)})\right\} ,\\
H_{\rho} & =(-1)^{\nu}I\left\{ ih\left[\frac{\sin2\alpha_{3-\nu}}{\sin2\alpha_{3}}B\cos[2(\varphi-\varphi_{\nu})]-\frac{1}{2}(-1)^{\nu}A+\frac{1}{2}B\right]r\right.\\
 & \hspace*{2.2cm}+\left.\frac{i\tau_{2}C}{k\sin(\tau_{2}\alpha_{\nu})}\cos[\tau_{2}(\varphi-\varphi_{\nu})]r^{\tau_{2}-1}+o(r^{\min(1,\tau_{2}-1)})\right\} ,\\
H_{\varphi} & =(-1)^{\nu}I\left\{ -\frac{ih\sin2\alpha_{3-\nu}}{\sin2\alpha_{3}}B\sin[2(\varphi-\varphi_{\nu})]r-\frac{i\tau_{2}C}{k\sin(\tau_{2}\alpha_{\nu})}\sin[\tau_{2}(\varphi-\varphi_{\nu})]r^{\tau_{2}-1}+o(r^{\min(1,\tau_{2}-1)})\right\} \\
j_{u} & =I\left\{ 2ihBr-\frac{C}{w}r^{\tau_{2}}+o(r^{\min(1,\tau_{2})})\right\} ,\\
j_{\rho} & =I\left\{ 2B+o(1)\right\} .
\end{split}
\label{eq:all}
\end{equation}
If $B=C=D=0$, then the field at the edge is described by smoother
functions compared with (\ref{eq:all}). Hereafter this trivial case
is not considered since it corresponds to a distribution of the field
of external sources that does not induce an electric current on the
RHP.

From the obtained expressions, it follows that $H_{u}$ and $j_{\rho}$
tend to constants when approaching the edge. The behavior of all other
components can be determined by comparing the two powers of $r$.
So, for example, for the component $E_{u}$ when $\alpha_{3}>\pi$,
the main term is proportional to $r^{\tau_{2}}$, while when $\alpha_{3}\leq\pi$,
the main term is proportional to $r$. In addition to the relationships
between angles, the behavior of the principal term can be influenced
by the value of the propagation constant $h$. If we take $h=0$,
i.e., consider a field that is constant along the $u$ axis, then
some terms in (\ref{eq:all}) will vanish, and the behavior of the
corresponding components will change. However, it can be noted that
the indicated change affects only those components that at $r\to0$
vanish no slower than linearly. The behavior of singular components,
or components not vanishing on the edge is not influenced by the value
of $h$. Just as in Refs.~\cite{Braver1986,Braver1988}, the obtained
expansions (\ref{eq:psi})--(\ref{eq:all}) do not allow one to carry
out a term-by-term limit transition $w\to\infty$ (the half-plane
is absent) or $w=0$ (the half-plane is perfectly conducting). In
each of these cases we arrive at the well-known \cite{Meixner1972,Mittra1971}
problem of determining the field near the edge of an perfectly conducting
wedge. The powers of $r$ that determine the order of the main term
in each component for different $h$ and $w$ are summarized in the
Table \ref{tab:1}, where it is noted that for $\alpha_{1}=\pi$ ($\alpha_{3}=\pi/2$)
the leading term of the transverse components of the electric (magnetic)
field cannot be described by a power function. The study of the field
at such values of the angles is carried out below. 

\begin{table}[H]
\caption{\label{tab:1}}

\centering{}%
\begin{tabular}{>{\centering}m{2cm}cccccc}
\toprule 
\addlinespace
\multirow{2}{2cm}{Field and current components} & \multicolumn{2}{c}{$w\neq0,\infty$} & \multicolumn{2}{c}{$w=0$} & \multicolumn{2}{c}{$w=\infty$}\tabularnewline\addlinespace
\cmidrule{2-7} \cmidrule{3-7} \cmidrule{4-7} \cmidrule{5-7} \cmidrule{6-7} \cmidrule{7-7} 
 & $h\neq0$ & $h=0$ & $h\neq0$ & $h=0$ & $h\neq0$ & $h=0$\tabularnewline
\midrule
\midrule 
\addlinespace
$H_{u}$ & 0 & 0 & 0 & 0 & 0 & 0\tabularnewline\addlinespace
\addlinespace
$E_{u}$ & $\min(1,\tau_{2})$ & $\tau_{2}$ & $\tau_{3}$ & $\tau_{3}$ & $\tau_{2}$ & $\tau_{2}$\tabularnewline\addlinespace
\addlinespace
$E_{\rho}$ & $\min(0,\tau_{3}-1)^{*}$ & $\min(0,\tau_{3}-1)^{*}$ & $\tau_{3}-1$ & $\tau_{3}-1$ & $\tau_{2}-1$ & $\tau_{2}-1$\tabularnewline\addlinespace
\addlinespace
$E_{\varphi}$ & $\min(0,\tau_{3}-1)^{*}$ & $\min(0,\tau_{3}-1)^{*}$ & $\min(1,\tau_{3}-1)$ & $\min(1,\tau_{3}-1)$ & $\min(1,\tau_{2}-1)$ & $\min(1,\tau_{2}-1)$\tabularnewline\addlinespace
\addlinespace
$H_{\rho}$ & $\min(1,\tau_{2}-1)^{**}$ & $\tau_{2}-1$ & $\min(1,\tau_{3}-1)$ & $\tau_{3}-1$ & $\min(1,\tau_{2}-1)$ & $\tau_{2}-1$\tabularnewline\addlinespace
\addlinespace
$H_{\varphi}$ & $\min(1,\tau_{2}-1)^{**}$ & $\tau_{2}-1$ & $\tau_{3}-1$ & $\tau_{3}-1$ & $\tau_{2}-1$ & $\tau_{2}-1$\tabularnewline\addlinespace
\addlinespace
$j_{u}$ & $\min(1,\tau_{2})$ & $\tau_{2}$ & $\min(1,\tau_{3}-1)$ & $\min(1,\tau_{3}-1)$ & --- & ---\tabularnewline\addlinespace
\addlinespace
$j_{\rho}$ & 0 & 0 & 0 & 0 & --- & ---\tabularnewline\addlinespace
\midrule 
\addlinespace
\begin{cellvarwidth}[m]$^{*}\alpha_{1}\neq\pi$

$^{**}\alpha_{3}\neq\pi/2$\end{cellvarwidth} &  &  &  &  &  & \tabularnewline
\end{tabular}
\end{table}

Let us now consider the degenerate cases, when the determinant $\Delta_{0}$
has roots $\tau_{s}$ and $\tau_{s'}>\tau_{s}$ that differ by an
integer $l$. Substituting the root $\tau_{s}$ into the determinant
$\Delta_{m}$, $m\geq1$, we get that at least one of them, namely
$\Delta_{l}$, is equal to zero. In this case, the appearance of a
term $r^{\tau_{s}+l}\ln r$ in the expansions of $\psi$ and $\chi$
is possible. The condition $\tau_{s'}-\tau_{s}=l$ is necessary but
not sufficient for the emergence of such a term. For example, for
$1/\alpha_{1}-1/\alpha_{3}=1/\pi$, despite the relation $\tau_{3}-\tau_{2}=1$,
the term proportional to $r^{\tau_{2}+1}\ln r$ does not appear in
the expansions of $\psi$ and $\chi$. This is due to the fact that
the solvability condition of the SLAE for integration constants appeaaring
in the expressions for $a_{10}$, $b_{10}$ at $\tau=\tau_{2}$ leads
to the equality $a_{11}=b_{11}\equiv0$. In the other case, when $1/\alpha_{2}-1/\alpha_{1}=1/\pi$,
the coefficients $a_{11}$ and $b_{11}$ in front of $r^{\tau_{3}+1}\ln r$
no longer turn to zero, and logarithmic functions remain in the series
(\ref{eq:psi-chi}). It is obvious that certain roots of the determinant
$\Delta_{0}$ differ by an integer also in the case when the ratio
of $\pi$ to any of the angles $\alpha_{1}$, $\alpha_{2}$, $\alpha_{3}$
is a rational number. The inconsistency of the Meixner method for
such angles in a structure consisting of dielectric wedges is shown
in the work \cite{Andersen1978}. The necessity of introducing logarithmic
terms for such a structure is shown in work \cite{Makarov1986}, where
it is noted that for dielectric structures the introduction of logarithmic
functions does not affect the first terms of the Meixner series, which
determine the behavior of the field at the edge.

\begin{table}[t]
\caption{\label{tab:2}}

\centering{}%
\begin{tabular}{>{\centering}m{2cm}lcc>{\centering}p{2cm}ccc}
\toprule 
\addlinespace
\multirow{2}{2cm}{Field and current components} & \multirow{2}{*}{$\alpha_{1}=\pi$} & \multicolumn{2}{c}{$\alpha_{3}=\pi/2$} & \multirow{2}{2cm}{Field and current components} & \multirow{2}{*}{$\alpha_{1}=\pi$} & \multicolumn{2}{c}{$\alpha_{3}=\pi/2$}\tabularnewline\addlinespace
\cmidrule{3-4} \cmidrule{4-4} \cmidrule{7-8} \cmidrule{8-8} 
 &  & $h\neq0$ & $h=0$ &  &  & $h\neq0$ & $h=0$\tabularnewline
\midrule
\addlinespace
$H_{u}$ & ${\cal O}(r^{0})$ & ${\cal O}(r^{0})$ & ${\cal O}(r^{0})$ & $H_{\rho}$, $H_{\varphi}$ & ${\cal O}(r^{\tau_{2}-1})$ & ${\cal O}(r\ln r)$ & ${\cal O}(r)$\tabularnewline\addlinespace
\addlinespace
$E_{u}$ & ${\cal O}(r^{\tau_{2}})$ & ${\cal O}(r)$ & ${\cal O}(r^{2})$ & $j_{u}$ & ${\cal O}(r^{\tau_{2}})$ & ${\cal O}(r)$ & ${\cal O}(r^{2})$\tabularnewline\addlinespace
\addlinespace
$E_{\rho}$, $E_{\varphi}$ & ${\cal O}(\ln r)$ & ${\cal O}(r^{0})$ & ${\cal O}(r^{0})$ & $j_{\rho}$ & ${\cal O}(r^{0})$ & ${\cal O}(r^{0})$ & ${\cal O}(r^{0})$\tabularnewline\addlinespace
\end{tabular}
\end{table}

In the structure under consideration, logarithmic functions may appear
in the leading terms of some field components. A simple analysis of
formulas (\ref{eq:all}) allows us to identify two such cases: $\alpha_{1}=\pi$
(Fig.~\ref{fig:1}b) and $\alpha_{3}=\pi/2$ (Fig.~\ref{fig:1}c).
The sequential solution of equations (\ref{eq:a''b''})--(\ref{eq:bc-2})
in each of these cases allows us to determine the behavior of all
field components. The results obtained for $w\neq0$, $w\neq\infty$
are summarized in Table \ref{tab:2}. Data for $w=0$ and $w=\infty$
is not provided since it may be found from Table \ref{tab:1} by substituting
the corresponding values of $\alpha_{1}$ or $\alpha_{3}$. In the
case $\alpha_{1}=\pi$ , there are no field components that vanish
linearly or faster when $r\to0$. For this reason, according to what
was mentioned above, the behavior the field at the edge at $\alpha_{1}=\pi$
does not depend on the values of $h$. If $\alpha_{2}=\alpha_{1}=\pi$,
then the wedge degenerates into a half-plane, and the data of Table
\ref{tab:2} turns into the corresponding results of work \cite{Braver1988}.
In the case of $\alpha_{3}=\pi/2$, the behavior of some field components
turns out to depend on the value of $h$. An interesting fact for
this structure is the disappearance at $h=0$ of the logarithmic functions
in the main terms of the transverse components of the magnetic field.

\begin{figure}[b]
\begin{centering}
\includegraphics{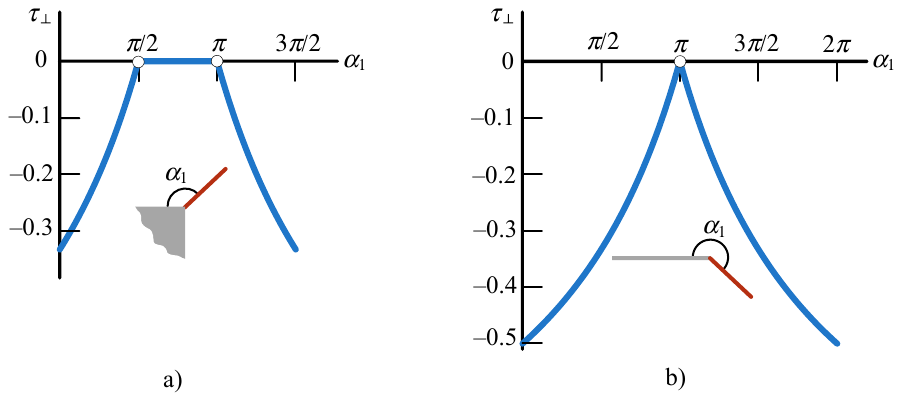}
\par\end{centering}
\caption{\label{fig:2}Singularity index $\tau_{\perp}$ of components $E_{\rho}$
and $E_{\varphi}$ as the function of $\alpha_{1}$. The holes in
the graphs correspond to $E_{\rho},E_{\varphi}={\cal O}(\ln r)$.}
\end{figure}

The set of data in Tables \ref{tab:1} and \ref{tab:2} determines
the behavior of the field near the common edge of a perfectly conducting
wedge and the RHP for all possible values of $w$, $h$, and angles
$\alpha_{1}$, $\alpha_{2}$. A similar structure is considered in
the work of Lang \cite{Lang1973}, where it is states that the presence
of an RHP always excludes the singularity of the transverse components
of the electric field. To test this statement, Fig.~\ref{fig:2}
shows the curves of singularity index $\tau_{\perp}$ of components
$E_{\rho}$, $E_{\varphi}={\cal O}(r^{\tau_{\perp}})$ for two structures
considered in Ref.~\cite{Lang1973}. When constructing these graphs,
for greater clarity, the condition $\alpha_{1}\geq\alpha_{2}$ used
above was omitted. From the provided graphs it follows that for a
structure with a rectangular wedge (Fig.~\ref{fig:2}a) the singularity
is excluded only at $\pi/2<\alpha_{1}<\pi$. For a structure with
a perfectly conducting half-plane (Fig.~\ref{fig:2}b), $E_{\rho}$
and $E_{\varphi}$ are singular for an arbitrary orientation of the
RHP.

The reason for the discrepancy between our results and Lang\textquoteright s
results \cite{Lang1973} is that the condition (B) (see Eq.~(5a)
in \cite{Lang1973}) is erroneous. The correct form of this formula
(in the notation of work \cite{Lang1973}) is as follows: $A_{0}^{(j)}=B_{0}^{(j)}=C_{0}^{(j)}=c_{0}^{(j)}=0$.
At the same time, $a_{0}^{(j)},b_{0}^{(j)}\neq0$ and the characteristic
equations (5b) and (5c) take the form $\sin\tau\varphi_{1}=0$ or
$\sin\tau(\varphi_{2}-\varphi_{1})=0$. In addition to the conditions
(A), (B) considered in \cite{Lang1973}, another condition, $A_{0}^{(j)}=B_{0}^{(j)}=c_{0}^{(j)}=0$,
is possible, under which $C_{0}^{(j)},a_{0}^{(j)},b_{0}^{(j)}\neq0$
and $\tau=1$. The existence of a solution corresponding to $\tau=1$,
for which all components of the field near the edge are finite, is
mentioned in the monograph \cite{Mittra1971}. Let us note here that
$\tau=1$ in the notations of works \cite{Mittra1971,Lang1973} corresponds
to $\tau=0$ in the notations of this work. Correct consideration
of all three conditions allows us to obtain, by Meixner method, the
results that coincide with the data in Table \ref{tab:1}. However,
degenerate cases (see Table \ref{tab:2}) can be investigated only
with the help of series (\ref{eq:psi-chi}) we introduced.

\section{Numerical results}

\begin{figure}[b]
\begin{centering}
\includegraphics{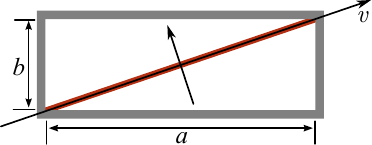}
\par\end{centering}
\caption{\label{fig:3}Cross-section of a rectangular waveguide with a resistive
film placed in the diagonal plane.}
\end{figure}

Let us demonstrate the efficiency of the found field distribution
near the edge by the example of the solution by the Galerkin method
of the integral equation \cite{Braver1984} for a surface current
density $\mathbf{j}$ on the resistive film in the diagonal plane
of the rectangular waveguide (see Fig.~\ref{fig:3}). Let us expand
$j_{u}$ and $j_{v}$ components ($u$ is the coordinate along the
waveguide axis) into the series of Gegenbauer\textquoteright s polynomials
$C_{\mu}^{(\tau)}(x)$ with the weight chosen according to the data
given in Table \ref{tab:2}:
\begin{equation}
\begin{split}j_{u} & =I\left[1-\left(\frac{2v}{d}\right)^{2}\right]\sum_{\mu=1}^{M^{u}}A_{\mu}C_{2\mu-1}^{(3/2)}\left(\frac{2v}{d}\right),\qquad d=\sqrt{a^{2}+b^{2}},\\
j_{v} & =I\sum_{\mu=1}^{M^{v}}B_{\mu}C_{2\mu-2}^{(1/2)}\left(\frac{2v}{d}\right).
\end{split}
\label{eq:Galerkin}
\end{equation}
The realization of the Galerkin method leads to a homogeneous system
of linear algebraic equations for determining the unknown coefficients
$A_{\mu}$, $B_{\nu}$. The condition of the existence of a nontrivial
solution of this system gives the required dispersion equation.

\vspace{1cm}

\pagebreak{}

Table \ref{tab:3} contains calculation results of the transversal
wavenumber $\kappa a$ of the dominant mode at $b/a=0.5$, $k_{0}a=5$,
$W=\unit[100]{\Omega}$, $\zeta_{0}=\unit[120\pi]{\Omega}$. Dominant
mode is the one which is transformed into the ${\rm TE}_{10}$ mode
of a hollow waveguide as $w\to\infty$. As we can see, already at
$M^{u}=2$, $M^{v}=3$ stabilization of six digits of $\kappa a$
is achieved. Note that using instead of (\ref{eq:Galerkin}) the corresponding
expansions of work \cite{Braver1984} for the same $M^{u}=2$, $M^{v}=3$
ensures the stabilization of only three significant digits of the
value of $\kappa a$. This is due to the fact that the basis functions
used in the work \cite{Braver1984} incorrectly describe the behavior
of the current near the edge.

\begin{table}
\caption{\label{tab:3}}

\centering{}%
\begin{tabular}{cccc}
\toprule 
\multirow{3}{*}{$M^{v}$} & \multicolumn{3}{c}{$M^{u}$}\tabularnewline
\cmidrule{2-4} \cmidrule{3-4} \cmidrule{4-4} 
 & 1 & 2 & 3\tabularnewline
\cmidrule{2-4} \cmidrule{3-4} \cmidrule{4-4} 
 & \multicolumn{3}{c}{$\kappa a$}\tabularnewline
\midrule
\addlinespace
1 & $3.59301-0.21609i$ & $3.59301-0.21609i$ & $3.59301-0.21609i$\tabularnewline\addlinespace
\addlinespace
2 & $3.61637-0.23759i$ & $3.61656-0,23757i$ & $3.61656-0.23757i$\tabularnewline\addlinespace
\addlinespace
3 & $3.61631-0.23771i$ & $3.61637-0.23771i$ & $3.61637-0.23771i$\tabularnewline\addlinespace
\addlinespace
4 & $3.61631-0.23771i$ & $3.61637-0.23771i$ & $3.61637-0.23771i$\tabularnewline\addlinespace
\end{tabular}
\end{table}

\end{document}